\documentclass[12pt,prd,aps,amssymb,amsmath,tightenlines]{revtex4}
\usepackage{graphicx}
\usepackage{dcolumn}
\usepackage{bm}
\usepackage[english]{babel}
\usepackage[latin1]{inputenc}
\usepackage{graphicx}
\usepackage{epstopdf}
\usepackage{amsfonts}
\usepackage{color}
\usepackage{epsfig}
\usepackage{mathrsfs}

\usepackage{amsmath,amssymb}
\usepackage{hyperref}

\usepackage{framed}
\usepackage[svgnames]{xcolor}

\renewcommand{\(}{\left(}
\renewcommand{\)}{\right)}
\renewcommand{\[}{\left[}
\renewcommand{\]}{\right]}

\newcommand{\be}{\begin{equation}}
\newcommand{\ee}{\end{equation}}

\date{}

\begin{document}

\selectlanguage{english}


\title{Simultaneous observation of gravitational and 
electromagnetic waves}

\author{Vincenzo Branchina}
\affiliation{Department of Physics and Astronomy, University of Catania, 
and INFN, Via Santa Sofia 64, I-95123 
Catania, Italy}

\author{Manlio De Domenico}
  \affiliation{Departament d'Enginyeria Inform\'atica i Matem\'atiques, Universitat Rovira i Virgili, Tarragona, Spain}

\date{\today}

\begin{abstract}
Assuming that the short gamma-ray burst detected by the Fermi Gamma-Ray
Space Telescope about 0.4~seconds after the gravitational waves observed by the
LIGO and VIRGO Collaborations originated from the same black hole merger event,
we perform a model-independent analysis of different quantum gravity scenarios
based on (modified) dispersion relations (typical of quantum gravity models) for the
graviton and the photon. We find that only scenarios where at least one of the
two particles is luminal (the other being sub- or super-luminal) are allowed, while
scenarios where none of the two particles is luminal are ruled out. Moreover, the
physical request of having acceptable values for the quantum gravity scale imposes
stringent bounds on the difference between the velocities of electromagnetic and
gravitational waves, much more stringent than any previously known bound.
\end{abstract}

\maketitle

\flushbottom


\section{Introduction}

Quantum gravity effects are usually expected to be relevant at energies 
around the Planck scale\,\cite{amelino1999gravity, amelino2001planck,  
amelino2002quantum,
bojowald2005loop,abdo2009limit,sotiriou2009phenomenologically,
liberati2011quantum,liberati2013tests}, 
$M_{Pl}\approx 1.2 \times 10^{19}$~GeV, where 
a breakdown of classical space-time in favor 
of a fuzzy/foamy description is usually expected\,\cite{amelino1998tests,
ashtekar2004background,woodard2009far,blas2010consistent,
vasileiou2015planck}. 
Obviously, quantum gravity theories cannot be 
tested at laboratory energies, but the propagation for cosmological distances 
of (ultra-) high energy particles  provides an 
excellent laboratory for testing deviations from classical general
relativity. There is great interest for the cases when multiple 
observations from the same astrophysical source are available and 
a time-of-flight analysis can be carried on~\cite
{coleman1999high,ng2003selected,jacobson2003threshold,carroll2003quantum,
amelino2004phenomenology,lamon2008glast,maccione2009planck,
scully2009lorentz,abdo2009limit,vasileiou2013constraints}, in particular
on using the observation of gravitational 
waves~\cite{mirshekari2012constraining} for this purposes.

Together with the discovery of the Higgs boson\,\cite{aad2012observation,chatrchyan2012observation} (and, if 
confirmed, of the $750$ GeV resonance\,\cite{atlas2016diphoton,cms2016diphoton}), a new great finding  provides another breakthrough, crucial for our understanding 
of fundamental physics and for the ways it opens for future research. 
This is the recent direct observation by LIGO and VIRGO Collaborations\,\cite{abbott2016} 
of gravitational waves generated by a black hole merger,
with a false alarm rate estimated to be less than 1 event per 203,000 years 
-- equivalent to a significance of 5.1$\sigma$ --, that called for enthusiastic 
multi-messenger searches from several observatories that tried to identify 
candidate photons and neutrinos subsequent to this GW150914 event. 
The gamma-ray observatories INTEGRAL~\cite{savchenko2016integral}, SWIFT~\cite{evans2016swift} and AGILE~\cite{tavani2016agile} did not observe significant photon excesses in different energy ranges. The neutrino observatories ANTARES and IceCUBE did not observe significant candidate neutrinos as well~\cite{adrian2016neutrinos}. At variance with other gamma-ray burst observatories, covering only a fraction of the sky where LIGO detected the GW150914 event, the Fermi Gamma-ray Space Telescope was exposed to a large fraction of the same region, and reported the detection of a weak gamma-ray burst above 50~keV just 0.4~s after GW150914, with a positional uncertainty region overlapping with that of the LIGO observation. The estimated false alarm probability for this observation is 0.0022~\cite{connaughton2016fermi}.

Whether or not the gravitational wave and 
the gamma-ray burst come from the same source, i.e whether the theory 
can accomodate both these signals from the merging of the two black 
holes is still a controversial question\,\cite{lyutikov2016fermi,loeb2016electromagnetic}. According 
to\,\cite{lyutikov2016fermi}, for instance, the two signals are unlikely 
to be related, and from the merging of these two black holes
no burst of photons should be observed. On the other hand, a recent 
model\,\cite{loeb2016electromagnetic} suggests that the merging 
black holes might have been generated by the collapse of a rapidly 
rotating massive star, that at the end of its collapse might have 
produced gamma-ray bursts. Other 
studies~\cite{li2016long,li2016implication,fraschetti2016possible,
perna2016short,yamazaki2016electromagnetic,janiuk2016gray} also try 
to reconcile the two observations, and this possible multi-messenger
signal has even been used to question the concept of cosmic 
acceleration from a genuine scalar-tensor modification of 
gravity~\cite{lombriser2016challenges}. 

In the following we assume that the gamma-ray burst observed by the Fermi Collaboration has been caused by the black hole merger that generated the gravitational waves detected by the LIGO Collaboration. As already noted, there is only a $0.2$ percent chance that these two events originated in the same patch of sky at the same time but were due to two different high-energy phenomena. Naturally, it is a challenge for present and future theoretical work to explain the mechanism that produces this (unexpected) electromagnetic signal in a black holes collision. Here we are interested in investigating the  consequences of this joint event, and we will see that they are very important. This issue has already been partly studied by a number of authors\,\cite{blas2016constraining,li2016implication,ellis2016comments,bonetti2016photon}
with different purposes, as deriving upper bounds for the 
speed of gravitational waves, for the graviton mass, or lower bounds for 
the quantum gravity scale. Actually, 
quantum gravitational effects usually modify the dispersion relation $E^{2}=p^{2}c^{2}+m^{2}c^{4}$ of a relativistic particle by terms that in an effective field theory approach depend on inverse powers of the quantum gravity scale  $E_{QG}$, and involve a violation of Lorentz invariance (LIV) at high energies, with $E_{QG}$ typically being of the order of the Planck scale $M_{Pl}\,c^2 \sim 10^{19}$ GeV. 

The aim of the present work is twofold. With the help of the modified dispersion relations for the graviton and the photon, on the one hand we study the upper bounds for the difference in speed between the gravitational and the electromagnetic wave, so to get in particular constraints on the gravitons' speed, on the other hand we perform a general model-independent analysis to constrain quantum gravity theories. We will find new and important results that greatly improve our knowledge on these  issues.   


The rest of the paper is organized as follows. In section II we 
consider the general set-up for our analysis, starting by 
considering the modified dispersion relation for massless particles 
and deriving the corresponding difference $\Delta t$ between 
the graviton and the photon propagation times, $\Delta t_g$ and 
$\Delta t_{\gamma}$ respectively. In section III we will capitalize on the theory to constraint the speed of gravitons, improving the current estimations by several orders of magnitude. In section IV we will fully develop the model-independent analysis to constrain both astrophysical models and quantum gravity theories, while section V is for our conclusions. 



\section{General setup}

Quantum gravity theories generically induce LIV, parametrized 
in terms of a model independent modified dispersion relation, that 
can be written as~\cite{vasileiou2013constraints} 
\begin{eqnarray}
\label{eq:disprel}E^{2} \approx p^{2}c^{2}
\times\[1+\sum_{s=1}^{\infty}\xi\(\frac{E}{E_{QG}}\)^{s}\],
\end{eqnarray}
where $E_{QG}$ is the quantum gravity scale, the factor $\xi$ accounts 
for sub-luminal ($\xi=-1$) and super-luminal ($\xi=+1$) propagation (while $\xi=0$ is for luminal particles),  
and $E \,\,(< E_{QG})$ is the energy of the particle. 

These 
corrections emerge for instance as leading-order terms in nonlinear 
quantum gravity models~\cite{amelino1999gravity}, and more generally 
they result from an effective field theory approach.
The linear correction accounts for violation 
of $\mathcal{C}\mathcal{P}\mathcal{T}$ symmetry, while when 
$\mathcal{C}\mathcal{P}\mathcal{T}$ symmetry is preserved, 
the linear term is absent and the 
dominant term is typically the quadratic correction.
Keeping in Eq.~(\ref{eq:disprel}) only the lowest-order 
non-vanishing term, say $s=n$, the particle (group) velocity is:
\begin{eqnarray}
\label{eq:vgroup}
v=\frac{\partial E}{\partial p}\approx 
c\times\[1+\xi\frac{n+1}{2}\(\frac{E}{E_{QG}}\)^{n}\]\,,
\end{eqnarray}
from which we see that when particles with different 
energies are produced by astrophysical objects at cosmological distances,
their observation by terrestrial detectors can be measurably 
delayed even if they are emitted at the same time. 

The idea of this work is to derive model-independent constraints on theories predicting LIV by considering the dispersion relation (\ref{eq:disprel}) for both photons and gravitons and applying the time-of-flight analysis to 
the recent observation of gravitational waves (gravitons) and the 
subsequent arrival of a gamma-ray burst (high-energy photons), under the assumption that both events have been caused by the same black hole merger event. 

To this end, in the following we use Eq.\,(\ref{eq:vgroup}) for photons and 
gravitons, considering all the possible cases obtained by combining 
$\xi_{\gamma}=-1,0,1$ with $\xi_{g}=-1,0,1$. In other words,
with no reference to any specific model, we allow 
for both gravitons and photons to be sub-luminal, luminal, or 
super-luminal, and making use of the experimental results, we 
analyse each of the corresponding combinations. 
Our analysis is performed under the very general assumption that 
that for both of them the dominant LIV contribution occurs for the same value of $n$. 

The difference $\Delta t=\Delta t_{g} -\Delta t_{\gamma}$ between the graviton and the photon propagation times is given by 
\be\label{deltat}
\Delta t=\Delta t_{a} -(1+z_0)\Delta t_{e}, 
\ee
where $\Delta t_{a}$ is the arrival delay observed at Earth and 
$\Delta t_{e}$ the emission delay at the source with redshift $z_0$, that in our case is $z_0=0.09$. 
While $\Delta t_{a}$ is a measured quantity, we do not have 
informations on $\Delta t_{e}$. The typical approach is to assume 
$\Delta t_{e}=0$, and to derive then bounds for the physical quantities 
of interest, for instance for the graviton velocity. 
In our analysis, we will allow $\Delta t_{e}$ to freely vary, and 
this will give us the possibility to investigate the compatibility of
different LIV models with the LIGO and Fermi data.

If $v_{\gamma}(z)$ and $v_{g}(z)$ are the speeds of the photons and 
the gravitons respectively at a given red-shift $z$, $\Delta t$ is given by:
\begin{eqnarray}\label{deltatt}
\Delta t=c H_{0}^{-1}\int_{0}^{z_0}\( \frac{1}{v_{g}(z)} - \frac{1}{v_{\gamma}(z)} \)
\frac{dz}{\sqrt{\Omega_{m}(1+z)^{3}+\Omega_{\Lambda}}},
\end{eqnarray}
where $\Omega_{m}$ is the fractional density of matter, $\Omega_{\Lambda}$ 
the fractional density of dark energy and $H_{0}$ the Hubble constant at 
present time in the $\Lambda-$CDM cosmological model. In the following, we will consider the values obtained from recent measures of the Planck collaboration~\cite{adam2015planck}, namely $\Omega_{m}\approx 0.31$, 
$\Omega_{\Lambda}\approx0.69$ and $H_{0}\approx67.8$~km/s/Mpc.

Inserting Eq.~(\ref{eq:vgroup}) for the energies $E_\gamma$ and 
$E_g$ of the photon and the graviton in Eq.\,(\ref{deltatt}) we get: 
\begin{eqnarray}
\label{eq:curves}
\Delta t\simeq \beta_{n}(z_0)\frac{n+1}{2}\[\xi_{\gamma}\(\frac{E_{\gamma}^0}{E_{QG}}\)^{n} - \xi_{g}\(\frac{E_{g}^0}{E_{QG}}\)^{n}\],
\end{eqnarray}
where $E_{g}^0$ and $E_{\gamma}^0$ are the energy of the graviton 
(gravitational wave) and of the photon (gamma-ray burst) meaured 
at Earth, and we have used 
$E_\gamma= h {\nu_\gamma}^0 (1+z) = E_\gamma^0 (1+z)$ and 
$E_g=h\nu_g^0 (1+z) = E_g^0 (1+z)$ (with $\nu_\gamma^0$
and $\nu_g^0$ the frequencies of the electromagnetic and gravitational
waves measured at Earth), and $\beta_{n}(z_0)$ is:
\begin{eqnarray}\label{betan}
\beta_{n}(z_0)&=&H_{0}^{-1}\int_{0}^{z_0}\frac{(1+z)^{n}dz}
{\sqrt{\Omega_{m}(1+z)^{3}+\Omega_{\Lambda}}}.
\end{eqnarray}

As we will see, Eq.\,(\ref{eq:curves}) is one of the key ingredients of our analysis. Other important ingredients are the following measured quantities: the time arrival delay $\Delta t_{a} \sim 0.41~s$ between the gamma-ray burst and the gravitational wave (see Eq.\,(\ref{deltat})) and the energies $E_{g}^0$ and $E_{\gamma}^0$ of the graviton and the photon observed at Earth. For the gamma-ray burst detected by the Fermi Gamma-Ray Burst Monitor (GBM), photons with energies between 50~keV and 1~MeV were observed 
\cite{connaughton2016fermi}. As for the gravitational wave, 
the signal sweeps upwards in frequency from 35 to 250~Hz, so that  
the energy of the gravitons at Earth is in the range $E_{g}^0=h\nu_{_0} \approx 
10^{-12} \,-\,10^{-13} $~eV, where $\nu_0$ is the gravitational wave 
frequency at Earth and $h$ is the Planck constant.

With no reference to any specific model, in the following 
we will explore different possible cases, namely we will consider that both the photon and the graviton can be super-luminal, luminal or sub-luminal. This means that for both particles we will consider in Eq.\,(\ref{eq:vgroup}) the following three different possibilities for $\xi$: $\xi=0, 1, -1$. This model-independent analysis will allow to use  Eq.\,(\ref{eq:curves}) to read the relation between the Quantum Gravity scale $E_{QG}$ and the difference $\Delta t_e$ (see Eq.\,(\ref{deltat})) in the emission time between the gravitational and the electromagnetic wave. One of the important results of this analysis will be that some models are ruled out simply by the fact that only too low values of the Quantum Gravity scale would be compatible with the observed values of  $\Delta t_{a}$, $E_{g}^0$ and $E_{\gamma}^0$. Before moving to that complete analysis (developed in section IV), in the following section we derive the constraints that Eq.\,(\ref{eq:curves}) puts on the speed of gravitational waves.    

\section{Constraining the speed of gravitational waves}

When multiple observations from the same astrophysical source are available, which is what we assume for the gravitational waves observed by the LIGO and VIRGO Collaborations and the gamma-ray burst observed by the Gamma-ray Space Telescope of the FERMI Collaboration, a typical time-of-flight analysis can be applied to derive an upper bound on the speed $v_{g}$ of gravitational waves. In the following we derive in this manner an upper bound on $v_{g}$ and show that this bound coincides (as it should be) with the one  derived in Ref.~\cite{ellis2016comments}.  
However, we will see that
the knowledge of the energies of the gravitational and the electromagnetic waves allows to put much more stringent upper bounds on $v_{g}$. This will be the first (in our opinion very important) of our new results. 

Actually, several bounds have already been derived in the literature. 
For instance, based on the direct observation of GW150914, an upper bound has been recently given by Blas \emph{et al}~\cite{blas2016constraining} and reads $v_{g}<1.87~c$.
Also, the tightest model-independent lower bound is $1-v_{g}/c\leq 2\times 10^{-15}$, deduced from the absence of gravitational Cherenkov radiation allowing for the unimpeded propagation of high-energy cosmic rays across our galaxy~\cite{moore2001lower}. A stricter (although model-dependent) bound from the same authors is $\approx 10^{-19}$, obtained under the assumption that the highest energy cosmic rays are produced by the so-called Z-burst mechanism. This mechanism predicts that very high energy neutrinos produced at cosmological distances annihilate with relic neutrinos via the Z-boson resonance~\cite{weiler1982resonant,weiler1999cosmic}, but it has been almost ruled out by recent limits on cosmic-ray photon fractions at the EeV energy scale~\cite{bleve2015limits}. Poorer bounds have been obtain from cosmology \cite{bellini2015constraints}, whereas $c/v_{g}-1 \leq 0.01$ has been deduced from radiation damping in binary systems~\cite{yagi2014strong} and observations of the Hulse-Taylor pulsar~\cite{jimenez2016evading}.

Let us see now how from Eq.\,(\ref{eq:vgroup}) we can derive an upper 
bound for the difference between the graviton and the photon speed 
from the typical time-of-flight analysis. Writing Eq.\,(\ref{eq:vgroup}) at Earth, we have:
\begin{eqnarray}\label{vminusv}
v_{\gamma}^0 - v_{g}^0= c\[
\xi_\gamma\frac{n+1}{2}
\(\frac{E_\gamma^0}{E_{QG}}\)^{n} -
\xi_g\frac{n+1}{2}
\(\frac{E_g^0}{E_{QG}}\)^{n}\]
\end{eqnarray}
where $v_\gamma^0$ , $v_g^0$ are the values of speed of the photon and the graviton measured at Earth ($E_\gamma^0$ and $E_g^0$, the energies of the photon and the graviton measured at Earth, have already been introduced before). Inserting then Eq.\,(\ref{eq:curves}) in Eq.\,(\ref{vminusv}) we have: 
\begin{eqnarray}\label{diffv}
v_{\gamma}^0- v_g^0
=\frac{c}{\beta_n(z_0)} \(\Delta t_{a} -(1+z_0)\Delta t_{e}\)
\end{eqnarray}
with $\beta_n(z_0)$ given in (\ref{betan}).

Assuming that the electromagnetic wave is emitted later than the gravitational wave, i.e. assuming that $\Delta t_{e} \geq 0$,  
we obtain 
the typical conservative limit by inserting $\Delta t_{e}=0$ in 
Eq.\,(\ref{diffv}), thus getting the upper bound 
\begin{eqnarray}\label{upperbound}
v_{\gamma}^0- v_g^0
\leq \frac{c}{\beta_n(z_0)} \Delta t_{a}
\end{eqnarray}

When the experimental values for $H_0$, $\Omega_M$, $\Omega_\Lambda$, $z_0$ and $\Delta t_a$ are used, we find
\begin{eqnarray}\label{upperbound1}
\frac{v_{\gamma}^0}{c}- \frac{v_g^0}{c}
\leq \,9.8 \cdot 10^{-18}\, \,\,\,\,\,\, {\rm for}\,\,\, n=1 
\end{eqnarray}
and 
\begin{eqnarray}\label{upperbound2}
\frac{v_{\gamma}^0}{c}- \frac{v_g^0}{c}
\leq \,9.4 \cdot 10^{-18}\,  \,\,\,\,\,\, {\rm for}\,\,\, n=2 
\end{eqnarray}
that are perfectly consistent with the bound found in Ref.~\cite{ellis2016comments}.

Up to now we have considered the usual time-of-flight analysis that brings  
to the upper bound found above for the difference between the photon and the 
graviton speed. 
However, the measured energies of the gravitons and photons that reach the terrestrial detectors, 
$E_{g}^0 \sim 10^{-12} \,-\,10^{-13}$ ~eV and $E_{\gamma}^0\,\sim\,50$~keV - $1$~MeV, allow to put much more interesting and stringent bounds on the speed of the gravitational waves. Moreover, we will see in the next section that using $\Delta t_{e} = 0$ to derive bounds on the graviton speed is not 
always consistent with the measured values of the photons and gravitons energies. In particular, for the case that is typically considered, and that we also investigate below, namely the case of a luminal photon and a sub-luminal graviton, $\Delta t_{e}$ cannot vanish unless the quantum gravity scale becomes (unacceptably) too low.    

Let us consider now the above mentioned case when the photon is luminal and the graviton is sub-luminal (or even super-luminal). From Eq.\,(\ref{vminusv}), considering the $n=1$ case and assuming that the Quantum Gravity scale $E_{QG}$ does not lie below the Planck scale, $E_{QG} \geq M_{Pl} c^{2}$, we have  
\begin{eqnarray}\label{fantastic}
\left|1 - \frac{v_{g}^0}{c}\right| \leq  10^{-40}\,.
\end{eqnarray} 

This is a fantastically much more stringent bound than the
one obtained in the corresponding ``usual'' Eq.\,(\ref{upperbound1}). 
Moreover, we see that even if we allow for a Quantum Gravity scale 
$E_{QG}$ down to the (too low) TeV scale, still the upper bound on the 
speed of  the gravitational wave is as high as 
\begin{eqnarray}
\left| 1 - \frac{v_{g}^0}{c}\right|\leq  10^{-24}\,,
\end{eqnarray} 
that is still much more stringent than the bound in 
Eq.\,(\ref{upperbound1}). 
These results greatly improve the existent upper bounds on the speed of gravitational waves.

\begin{figure}[!t]
\centering
\includegraphics[width=0.95\textwidth]{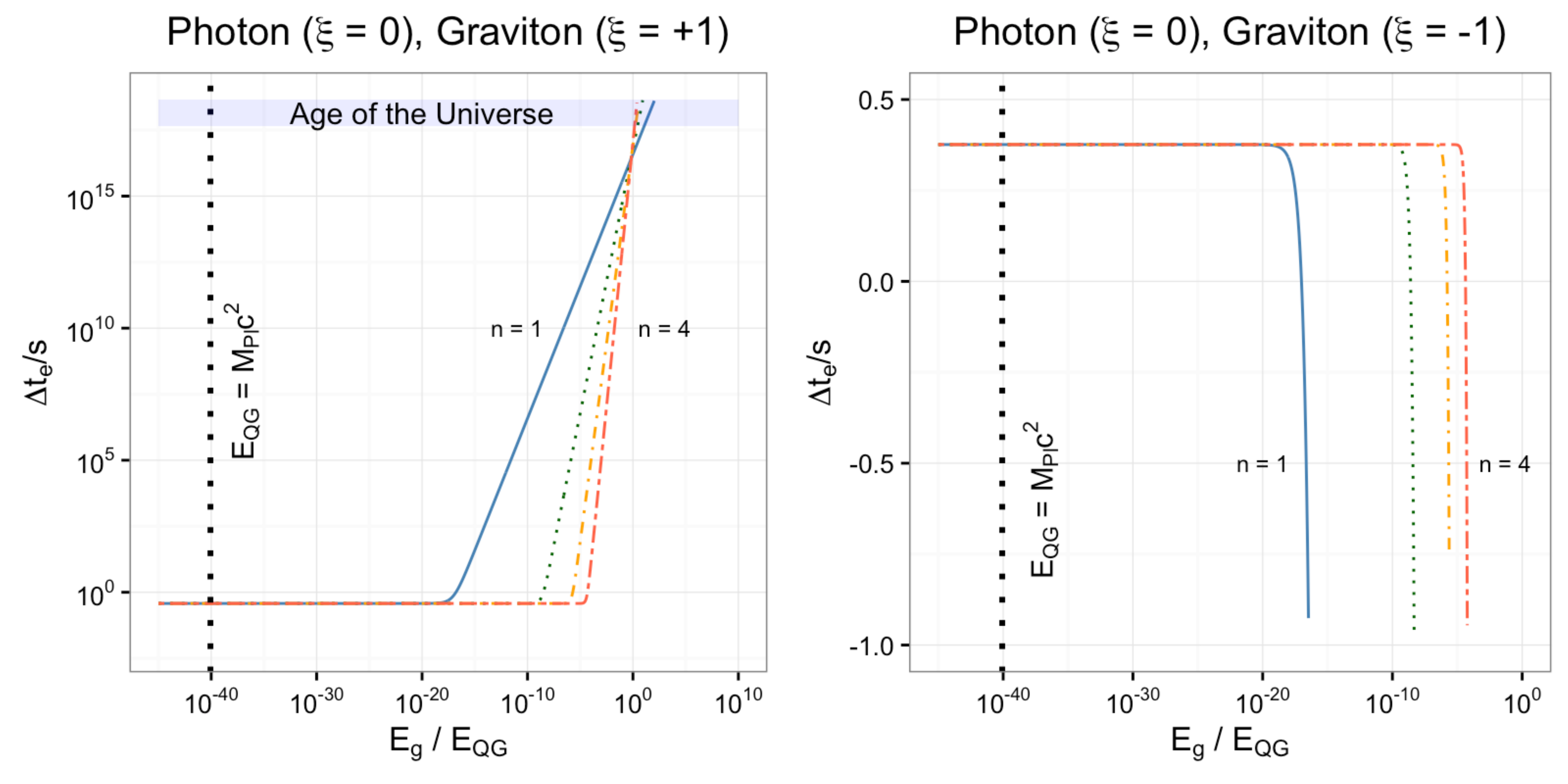}
\caption{Theoretical curves relating the time emission delay 
$\Delta t_{e}$ (between the gravitational wave and the gamma-ray burst) and the energy ratio $E_{g}^{0}/E_{QG}$ (where $E_{g}^{0}$ is the measured energy of the gravitons and $E_{QG}$ the quantum gravity scale) for the cases B (left panel) and C (right panel) of Tab.\,\ref{tab:models}, and for different values of $n$ ($n=1,2,3$ and 4), are shown. The gravitons and the photons obey to (quantum gravitationally modified) dispersion relations as in Eq.\,(\ref{eq:disprel}) (see also Eq.\,(\ref{eq:vgroup})). The dotted thick vertical line indicates the value of $E_{g}^{0}/E_{QG}$ obtained for $E_{QG}=M_{Pl}c^{2}$.}
\label{fig:bounds2-3}
\end{figure}

We can also consider the opposite case, when the photons are  sub- or super-luminal. From Eq.\,(\ref{vminusv}) we find that for $E_{QG} \geq M_{Pl} c^{2}$ the bound is
\begin{eqnarray}
|v_{\gamma}^0 - v_{g}^0| \leq \,c\,\cdot\, 10^{-22}\,,
\end{eqnarray}
that is a very important upper bound for the cases (rarely considered in the literature) of a sub- or super-luminal photon.

\section{Constraining Quantum Gravity Models}

Carrying on with our analysis, and with no reference to any specific model, we use now our general, model-independent Eq.\,(\ref{eq:curves}) to consider all possible scenarios involving sub-, super- and luminal photons and gravitons. Each scenario is governed by a different phenomenological equation, derived from Eq.\,(\ref{eq:curves}) and reported in Tab.~\ref{tab:models}, that provides the expected relationship between the emission time delay $\Delta t_{e}$ and the quantum gravity scale $E_{QG}$ for different cases ($n=1,2,3$ and 4) satisfying dispersion relations as in 
Eq.\,(\ref{eq:disprel}).

When both the photons and the gravitons are luminal, that is the case $A$ of 
Tab.~\ref{tab:models}, from Eqs.\,(\ref{deltat}) and (\ref{eq:curves}) we immediately have that the time delay $\Delta t_e$ between the emission of the gravitational wave and the subsequent emission of the gamma-ray burst is 
$\Delta t_{e}=\Delta t_{a}/(1+z_0)\sim 0.4 \, s$. In other words, we find  the obvious result that if both particles travel at the speed of light, the emission time and the arrival time are the same, up to the red-shift correction factor. Let us move now to consider the other less trivial cases. 

Fig.\,\ref{fig:bounds2-3} shows the phenomenological curves, for different values of $n$, relating $\Delta t_{e}$ and $E_{g}^{0}/E_{QG}$ for the cases B and C of Tab.\,\ref{tab:models}, where the photon  is luminal and the graviton is super-luminal (left panel) or sub-luminal (right panel). For any given value of $E_{QG}$ (and for any given value of $n$), the allowed value of $\Delta t_{e}$ is the one such that the theoretical curve intersects the corresponding vertical line. For instance, the dotted thick vertical
line of the figure indicates the value of $E_g^{0}/E_{QG}$ obtained for $E_{QG} = M_{Pl} c^2$. Note that $E_g^{0}$ is the value of the graviton energy measured at Earth, $E_g^0 \sim 10^{-12} - 10^{-13}$ eV, so that the curves actually provide a relation between the emission time delay $\Delta t_{e}$ and the quantum gravity scale $E_{QG}$.

\begin{large}
\begin{table}[!ht]
\begin{tabular}{cccl}
\hline\\
Case & $\xi_{\gamma}$ & $\xi_{g}$ & Equation\\\hline
A & 0 & 0 & $\Delta t_{e}=\Delta t_{a}/(1+z_{0})$\\
B & 0 & 1 & $\tilde{E}_{g}=[-\Delta t'_{a} + (1+z_{0})\Delta t'_{e}]^{1/n}$\\
C & 0 & -1 & $\tilde{E}_{g}=[\Delta t'_{a} - (1+z_{0})\Delta t'_{e}]^{1/n}$\\
D & 1 & 0 & $\tilde{E}_{\gamma}=[\Delta t'_{a} - (1+z_{0})\Delta t'_{e}]^{1/n}$\\
E & 1 & 1 & $\tilde{E}_{\gamma}=[\Delta t'_{a} - (1+z_{0})\Delta t'_{e} + \tilde{E}_{g}^{n}]^{1/n}$\\
F & 1 & -1 &$\tilde{E}_{\gamma}=[\Delta t'_{a} - (1+z_{0})\Delta t'_{e} - \tilde{E}_{g}^{n}]^{1/n}$\\
G & -1 & 0 & $\tilde{E}_{\gamma}=[-\Delta t'_{a} + (1+z_{0})\Delta t'_{e}]^{1/n}$\\
H & -1 & 1 & $\tilde{E}_{\gamma}=[-\Delta t'_{a} + (1+z_{0})\Delta t'_{e} - \tilde{E}_{g}^{n}]^{1/n}$\\
I & -1 & -1 & $\tilde{E}_{\gamma}=[-\Delta t'_{a} + (1+z_{0})\Delta t'_{e} + \tilde{E}_{g}^{n}]^{1/n}$\\
\hline
\end{tabular}
\caption{Phenomenological equations corresponding to models with sub-, super- or luminal photons and gravitons. Note that $\tilde{E}_{\gamma}=E_{\gamma}^{0}/E_{QG}$, $\tilde{E}_{g}=E_{g}^{0}/E_{QG}$, $\Delta t'_{a}=\Delta t_{a}/(\beta_{n}\frac{n+1}{2})$ and $\Delta t'_{e}=\Delta t_{e}/(\beta_{n}\frac{n+1}{2})$.}
\label{tab:models}
\end{table}
\end{large}
  
From this figure we see that the time delay  $\Delta t_{e}$ is practically quenched to the value $\Delta t_{e} \sim \Delta t_{a}/(1+z_0)\approx0.4$~s (that is the value obtained for the case A), for $E_{QG}>100~$keV in the $n=1$ case, and $E_{QG}>100~\mu$eV for the $n=2$ case, both values being several orders of magnitudes lower than the most stringent lower bounds currently available~\cite{vasileiou2013constraints} ($E_{QG}>7.6~M_{Pl}c^{2}$, for $n=1$, and $E_{QG}>10^{-9}~M_{Pl}c^{2}$, for $n=2$). We than see that acceptable values for the quantum gravity scale $E_{QG}$
are obtained only if $\Delta t_{e} \sim 0.4$\,s. 

As a consequence, if the photon is luminal and the graviton is super-luminal,
that is the case considered in the left panel of Fig.~\ref{fig:bounds2-3}, it is {\it never possible} to have $\Delta t_{e} = 0$, the latter being the condition needed to establish the upper bound in Eq.\,(\ref{upperbound1}) (or Eq.\,(\ref{upperbound2})). Actually, the complete analysis presented above shows that the finding 
$\Delta t_{e} \sim \Delta t_{a}/(1+z_{0}) \approx 0.4$~s, that in turn means 
$\Delta t \sim 0$, explains why we have obtained such a stringent bound as the one of Eq.\,(\ref{fantastic}), by far much more stringent than the bound of Eq.~(\ref{upperbound1}).
  
In the right panel of  Fig.~\ref{fig:bounds2-3} we consider the case when the photon is luminal and the graviton is sub-luminal. This is the case mostly studied in the literature, and in particular it contemplates the case when the photon is  luminal and the graviton is massive~\cite{will1998bounding}. 
As we can see from the theoretical curves shown in this figure, in this case  the value $\Delta t_e \sim 0$ can be reached, but this occurs for a value of $E_{QG}$ that for $n=1$ is more than $20$ orders of magnitudes below the Planck scale, and for greater values of $n$ is even smaller. As for the previous case, we then see that acceptable values for the quantum gravity scale $E_{QG}$ are obtained only if the time emission delay is frozen to the value $\Delta t_{e} \sim 0.4$\,s, and this again  provides an explanation for the extraordinarily low value of the upper bound $|1 - v^0_g/c| \leq 10^{-{40}}$ \, of Eq.\,(\ref{fantastic}).

\begin{figure}[!t]
\centering
\includegraphics[width=0.95\textwidth]{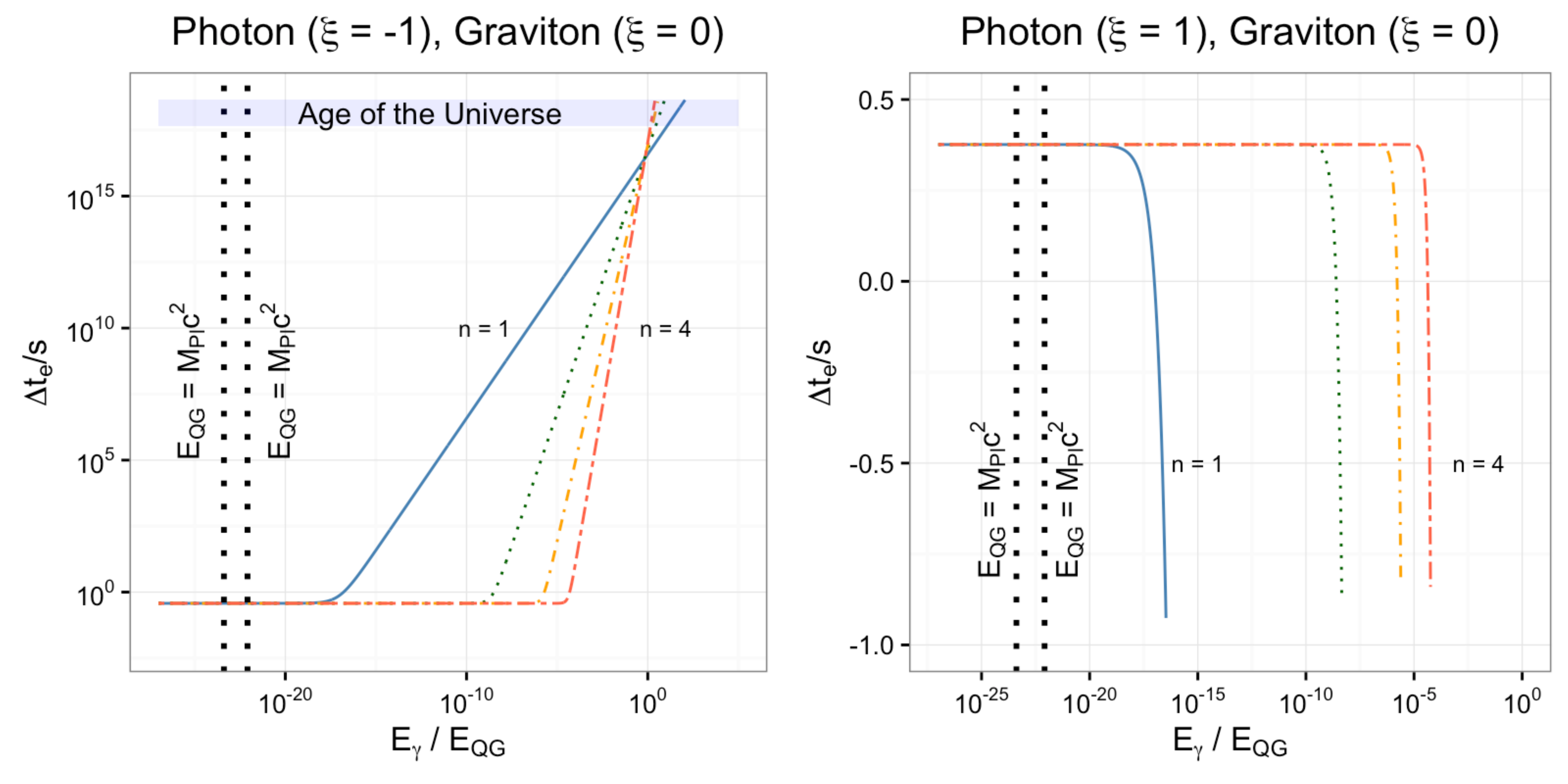}
\caption{Similarly to Fig.\,\ref{fig:bounds2-3}, here we plot the curves corresponding to the cases D (left panel) and G (right panel) of Tab.\,\ref{tab:models}. The dotted thick vertical lines indicate the values of $E_{\gamma}^{0}/E_{QG}$ obtained for $E_{QG}=M_{Pl}c^{2}$ and the extremal energies of the observed photons, 50 keV and 1 MeV. }
\label{fig:bounds4-7}
\end{figure}

Fig.\,\ref{fig:bounds4-7} shows the cases D and G of Tab.\,\ref{tab:models}, where the graviton is luminal and the photon sub- (left panel) or super-luminal (right panel). Similarly to the case of Fig.\,\ref{fig:bounds2-3}, here we present the curves that relate the time emission delay $\Delta t_{e}$ between the two waves with the energy ratio $E_{\gamma}^{0}/E_{QG}$, where $E_{\gamma}^{0}$ is the measured energy of the photons, for different values of $n$ ($n=1,2,3$ and 4). As for the previous cases, 
for any value of $E_{QG}$ the allowed value of $\Delta t_{e}$ is 
obtained from the intersection between the curve and the vertical line
corresponding to a given value of  $E_{\gamma}^{0}/E_{QG}$. For instance, the two dotted thick vertical lines shown in the figure indicate the values of $E_{\gamma}^{0}/E_{QG}$ obtained for $E_{QG} = M_{Pl} c^2$
and with the two extremal values of the energies of the observed photons, 
$50$ keV and $1$ MeV.  
Even for these cases we find that $\Delta t_{e}=\Delta t_{a}/(1+z_{0})\approx0.4$~s is the only allowed value for $E_{QG}>100~$GeV ($n=1$) and $E_{QG}>0.1~$eV ($n=2$), again much below the most stringent lower bounds~\cite{vasileiou2013constraints}.

\begin{figure}[!t]
\centering
\includegraphics[width=0.95\textwidth]{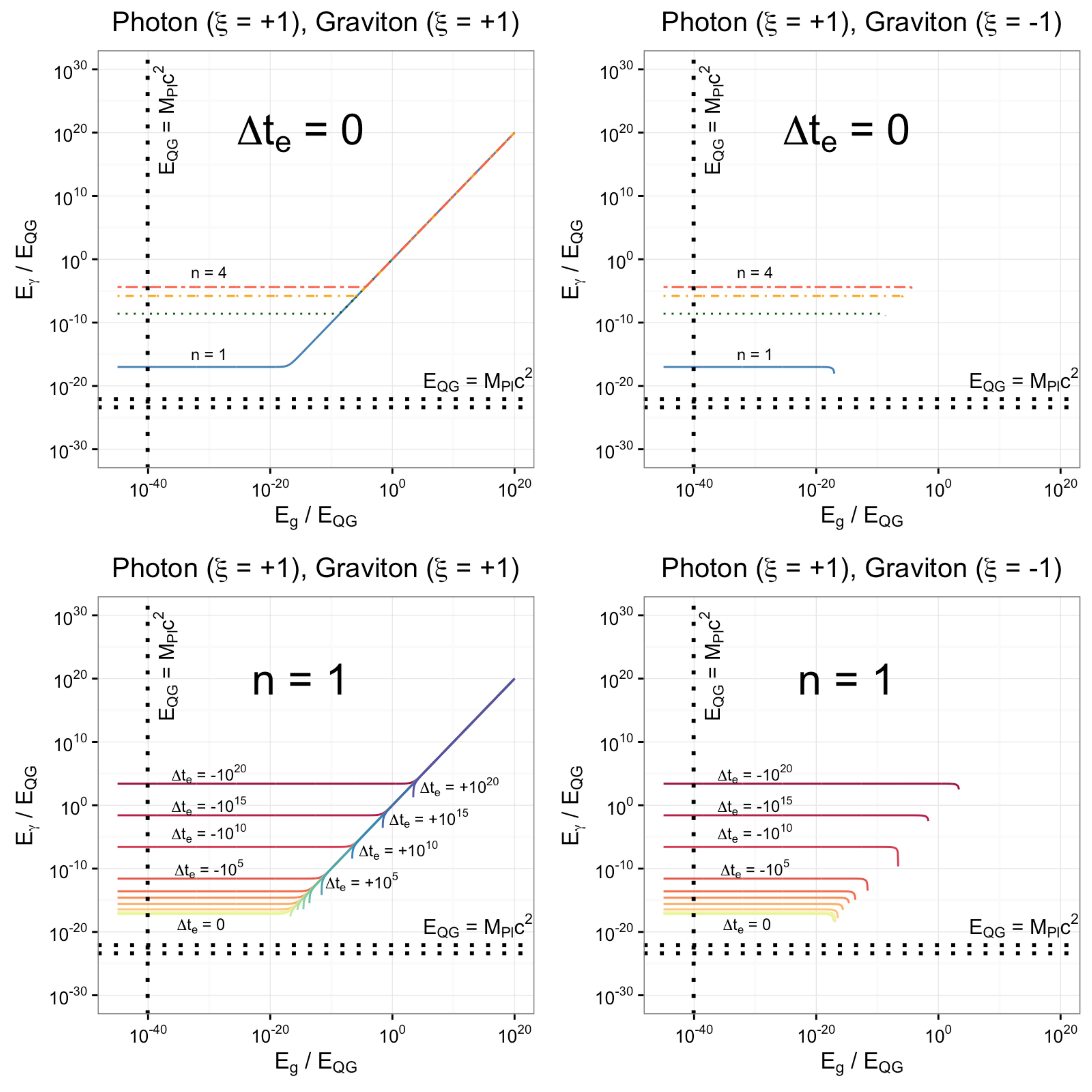}
\caption{The theoretical curves corresponding to the cases E (left panels) and F (right panels) of Tab.~\ref{tab:models} are plotted. As for the other figures, dotted thick lines correspond to the observed values of $E_g^{0}$ and $E_{\gamma}^{0}$ for $E_{QG}=M_{Pl}c^{2}$. In top panels, we consider $\Delta t_{e}=0$ and vary $n$, whereas in bottom panels we consider $n=1$ and vary $\Delta t_{e}$.
}
\label{fig:bounds5-6}
\end{figure}

From the analysis developed so far we can already draw some important lessons. First of all we note that scenarios where one of the particles (the photon or the graviton) is luminal and the other sub- or super- luminal are allowed, but the requirement of having an acceptable value for the quantum gravity scale imposes that the time emission delay between the gravitational and the electromagnetic waves has to be equal to $\Delta t_{a}/(1+z_{0})\approx 0.4$~s. This is an important result as it strongly constrains astrophysical models~\cite{loeb2016electromagnetic,li2016long,li2016implication,fraschetti2016possible,
perna2016short,yamazaki2016electromagnetic,janiuk2016gray} that aim to explain the arrival delay of the two signals and the production of short gamma-ray bursts from black hole mergers.
Moreover, the fact that $\Delta t_e \sim 0.4$ s, that in turn means $\Delta t \sim 0$, implies very stringent lower bounds on the difference in speed between the photon and the graviton, much more stringent than previous existing bounds, and this is in fact what we have found in the previous section.

Up to now we have considered cases where at least one of the two particles
(the photon or the graviton) is luminal. Now we move to consider cases when they are both super- or sub- luminal, or one of them is super- and the other is sub- luminal. The corresponding relationships between the physical quantities $E_{g}^{0}/E_{QG}$, $E_{\gamma}^{0}/E_{QG}$ and $\Delta t_e$ are obtained from Eq.\,(\ref{eq:curves}) and listed as cases E, F, H, I in Tab.\,\ref{tab:models}.

\begin{figure}[!t]
\centering
\includegraphics[width=0.95\textwidth]{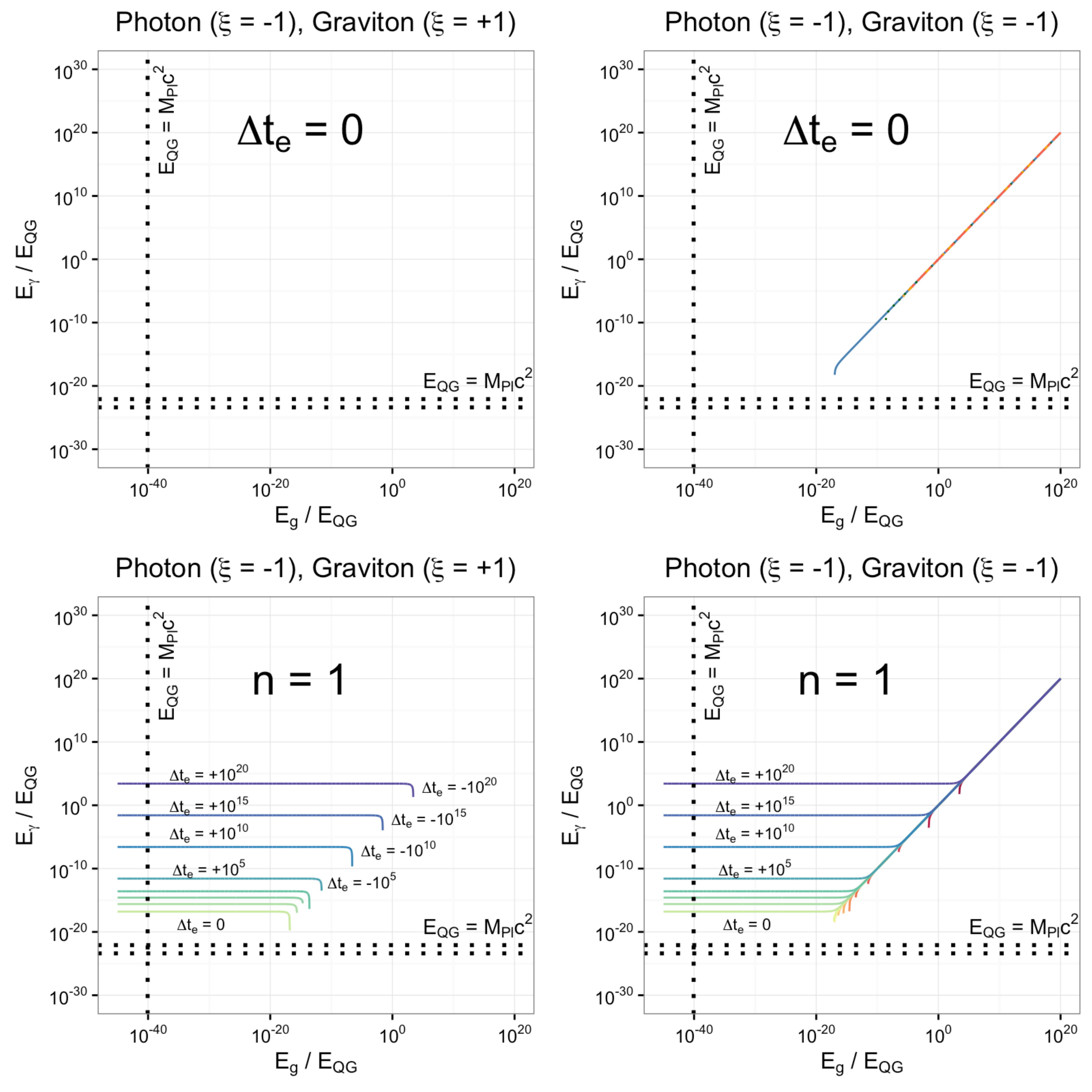}
\caption{As in Fig.~\ref{fig:bounds5-6}, here we consider the cases corresponding to models H (left panels) and I (right panels) of 
Tab.~\ref{tab:models}. In the top-left panel there are no curves because for $\Delta t_{e}=0$ the corresponding values of $E_{\gamma}^{0}/E_{QG}$ would be negative.}
\label{fig:bounds8-9}
\end{figure}

Results obtained for the case E and F, where the photon is super-luminal and the graviton is sub- (left panels) or super- (right panels) luminal are shown in Fig.\,\ref{fig:bounds5-6}. In the two top panels we consider the special case when there is no emission delay between the gravitational and the electromagnetic wave, i.e. when $\Delta t_{e}=0$. Under this condition, we are left with a relation between the two ratios $E_{\gamma}^{0}/E_{QG}$ and $E_{g}^{0}/E_{QG}$. As in the previous figures, the thick dotted lines are obtained for the measured values of $E_{g}^{0}$ and $E_{\gamma}^{0}$ when $E_{QG}=M_{Pl}\,c^2$. We see that the theoretical curves, even in the $n=1$ case, do not fit in the experimental allowed range, and only if the quantum gravity scale is lowered of several orders of magnitudes (so that the orizontal thick dotted lines move upwards, while the vertical ones move to the right) the curves become compatible with experiments. 

In the two other panels of Fig.\,\ref{fig:bounds5-6} we plot again the theoretical curves relating $E_{\gamma}^{0}/E_{QG}$ with $E_{g}^{0}/E_{QG}$ for several different values of $\Delta t_{e}$, ranging from
$\Delta t_{e}=10^{-20}$~s to $\Delta t_{e}=10^{20}$~s, considering only the $n=1$ case. The results are similar to those discussed above for the 
$\Delta t_e=0$ case. We see that there are no values of $\Delta t_{e}$ that are compatible at the same time with the measured value of $E_g^{0}$ and $E_{\gamma}^{0}$  and with a quantum gravity scale that lies at or above the Planck scale. Only if $E_{QG} \ll M_{Pl}c^{2}$ the theoretical curves would lie in the allowed region. For instance, in the case $E_{QG} \approx 10^{13}~$GeV ($n=1$) emission delays between -1~s and 0~s would be possible.  



Similar conclusions are drawn from the analysis of models H and I of 
Tab.\,\ref{tab:models}, and the corresponding theoretical curves are shown in Fig.~\ref{fig:bounds8-9}. 
Again, there are no values of $\Delta t_{e}$ compatible with observation and a quantum gravity energy scale above the Planck's one. Note that there are no curves in the top-left panel because for $\Delta t_{e}=0$ the corresponding values of $E_{\gamma}^{0}/E_{QG}$ would be negative. As for the previous cases, only if $E_{QG} \ll M_{Pl}c^{2}$ the theoretical curves would reach the allowed region. For instance, in the case $E_{QG} \approx 10^{13}~$GeV ($n=1$) emission delays between 0.4~s and 1~s would be allowed. Once again, these kind of models would be allowed only if the quantum gravity scale is downshifted much below the Planck scale. 

Therefore, from the analysis of the cases E F H I of Tab.\,\ref{tab:models} we learn the following very important lesson: scenarios where none of the two particles, the photon and the graviton, is luminal are not allowed, unless the quantum gravity scale lies in an energy range that is several orders of magnitude below the Planck scale.





\section{Conclusions}

Let us summarize the results of the present work. Assuming that the gamma-ray burst observed by the Fermi Collaboration was emitted by the black hole merger that produced the gravitational waves detected by the LIGO Collaboration, 
we have performed a model-independent analysis that allows to obtain bounds for the difference in speed between the gravitational and the electromafgnetic waves and constraints on quantum gravity models.

With the help of dispersion relations typical of many quantum gravity models, we have found that, when the photon is luminal while the graviton is sub- or super-luminal, if (as expected) the quantum gravity scale lies at or above the Planck scale, an extraordinarily stringent constrain on the speed of the gravitational wave emerges, namely $|1 - v_{g}^0/c| \leq 10^{-40}$.  This is a much more stringent limit than those obtained so far\,\cite{will1998bounding,ellis2016comments,li2016implication}.
Even if we lower considerably the quantum gravity scale, the upper bound on the speed of gravitational waves stays may orders of magnitudes lower than any previously known bound. Moreover, 
we have also considered the case where the photons are not luminal. In this case, always considering that $E_{QG}\geq M_{Pl}\,c^2$, we have constrained the difference in speed as $|v_{\gamma}^0 - v_{g}^0| \leq c\,\times \,10^{-22}$, that is still five orders of magnitude more stringent than those obtained with other approaches.

Our model-independent analysis also demonstrates that only scenarios with at least one luminal particle -- either the photon or the graviton or both -- are allowed, regardless of the value of $n$ in the modified dispersion relation. Other scenarios with no luminal particles are ruled out, unless the energy scale of quantum gravity is downshifted by several orders of magnitude. In all likely scenarios, the allowed values for the time emission delay are strictly constrained to equal $\Delta t_{a}/(1+z_{0})\approx 0.4$~s, result that strongly constrains the new astrophysical models~\cite{loeb2016electromagnetic,li2016long,li2016implication,fraschetti2016possible,perna2016short,yamazaki2016electromagnetic} (see Ref.~\cite{li2016long} for the description of some possible models and the corresponding emission delay) required to explain the production of short gamma-ray bursts from black hole mergers.



\clearpage
\newpage


%

\end{document}